\documentclass[11pt]{article}

\usepackage[margin=1.1in]{geometry}
\usepackage{graphicx}

\newcommand{\bfb}{{\bf b}}

\newcommand{\bfe}{{\bf e}}

\newcommand{\bfv}{{\bf v}}

\newcommand{\x}{{\bf x}}
\newcommand{\z}{{\bf z}}

\usepackage{amssymb, amsmath, amsthm}
\usepackage{thmtools}

\begin{document}

\title{Least Squares on GPUs in Multiple Double Precision\thanks{Supported
by the National Science Foundation under grant DMS 1854513.}}

\author{Jan Verschelde\thanks{University of Illinois at Chicago,
Department of Mathematics, Statistics, and Computer Science,
851 S. Morgan St. (m/c 249), Chicago, IL 60607-7045
Email: {\tt janv@uic.edu}, URL: {\tt http://www.math.uic.edu/$\sim$jan}.}}

\date{20 March 2022}


\maketitle

\begin{abstract}
This paper describes the application of the code generated by the
CAMPARY software to accelerate the solving of linear systems in
the least squares sense on Graphics Processing Units (GPUs),
in double double, quad double, and octo double precision.
The goal is to use accelerators to offset the cost overhead 
caused by multiple double precision arithmetic.
For the blocked Householder QR and the back substitution,
of interest are those dimensions at which teraflop performance is attained.
The other interesting question is the cost overhead factor that
appears each time the precision is doubled.

Experimental results are reported on five different NVIDIA GPUs,
with a particular focus on the P100 and the V100,
both capable of teraflop performance.
Thanks to the high Compute to Global Memory Access (CGMA) ratios
of multiple double arithmetic, teraflop performance is already attained
running the double double QR on 1,024-by-1,024 matrices,
both on the P100 and the V100.
For the back substitution, the dimension of the upper triangular system
must be as high as 17,920 to reach one teraflops on the V100,
in quad double precision, and then taking only the times spent by
the kernels into account.
The lower performance of the back substitution in small dimensions
does not prevent teraflop performance of the solver at dimension 1,024,
as the time for the QR decomposition dominates.

In doubling the precision from double double to quad double
and from quad double to octo double, the observed cost overhead factors
are lower than the factors predicted by the arithmetical operation counts.
This observation correlates with the increased performance for
increased precision, which can again be explained by the high CGMA ratios.
\end{abstract}


\section{Introduction}

Many applications in scientific computing may benefit from
extended precision, e.g.,~\cite{HD01} applies double doubles.
However, the cost overhead caused
by multiprecision arithmetic is a valid concern.
This paper experimentally demonstrates
that this cost overhead can be mitigated by the acceleration 
on a Graphics Processing Unit (GPU)
capable of teraflop performance.

The least squares solution $\x$ of a linear system $A \x = \bfb$ minimizes
the sum of the squares of~$\bfb - A \x$, or~$\| \bfb - A \x \|_2^2$.
The decomposition of the matrix~$A$ into an orthogonal matrix~$Q$
and an upper triangular matrix~$R$, $A = Q R$ reduces $A \x = \bfb$
to $R \x = Q^T \bfb$, solved by back substitution.
The Householder QR factorization
is numerically stable~\cite[Theorem~3.5]{Dem97}.

The blocked Householder QR factorization~\cite{BV87} 
is rich in matrix-matrix products~\cite{GV96},
well suited for GPU acceleration,
as demonstrated in~\cite{BDT08} and~\cite{VD08},
with further developments in~\cite{AADF11}, \cite{ABDK11}, 
\cite{KNDD12}, \cite{TDB10}, and~\cite{TNLD10}.
The development of the code for this paper benefited greatly
from the exposition in~\cite{KCR09}.
To develop a GPU accelerated back substitution algorithm,
ideas were taken from~\cite{NM01},
based on formulas proposed in~\cite{Hel78}. 

The suitability of double double and triple precision
Basic Linear Algebra Subroutines (BLAS) was shown in~\cite{MT12a,MT12b}.

\subsection{Multiple Double Arithmetic}

A multiple double number is an unevaluated sum of multiple doubles.
The arithmetical operations on multiple double numbers are defined
by algorithms in double precision arithmetic.
To extend the double precision $m$ times, compute with $m$ doubles.
Table~\ref{tabopscount} tallies the cost overhead to multiply
the precision with 2, 4, and 8, corresponding respectively 
to double double, quad double, and octo double precision,
to about 32, 64, and~128 decimal places of precision.
The averages (37.7, 439.3, 2379.0) predict the arithmetical cost
overhead factors.

\begin{table}[htb]
\begin{center}
\begin{tabular}{c|rrrrr}
& \multicolumn{5}{c}{double double: 37.7x} \\
& \multicolumn{1}{c}{$+$} & \multicolumn{1}{c}{$-$}
& \multicolumn{1}{c}{$*$} & \multicolumn{1}{c}{$/$}
& \multicolumn{1}{c}{$\Sigma$} \\ \hline
add &   8 &   12 &    &   & 20 \\
mul &   5 &    9 &  9 &   & 23 \\
div &  33 &   18 & 16 & 3 & 70 \\ \hline \hline
& \multicolumn{5}{c}{quad double: 439.3x} \\
& \multicolumn{1}{c}{$+$} & \multicolumn{1}{c}{$-$}
& \multicolumn{1}{c}{$*$} & \multicolumn{1}{c}{$/$}
& \multicolumn{1}{c}{$\Sigma$} \\ \hline
add &  35 &   54 &     &   &  89 \\
mul &  99 &  164 &  73 &   & 336 \\
div & 266 &  510 & 112 & 5 & 893 \\ \hline \hline
& \multicolumn{5}{c}{octo double: 2379.0x} \\
& \multicolumn{1}{c}{$+$} & \multicolumn{1}{c}{$-$}
& \multicolumn{1}{c}{$*$} & \multicolumn{1}{c}{$/$} 
& \multicolumn{1}{c}{$\Sigma$} \\ \hline
add &   95 &  174 &     &   &  269 \\
mul &  529 &  954 & 259 &   & 1742 \\
div & 1599 & 3070 & 448 & 9 & 5126
\end{tabular}
\caption{Operational counts for double double, quad double,
and octo double arithmetic.  For example:
one division (div) of two quad doubles requires
266 additions ($+$), 510 subtractions ($-$), 112 multiplications ($*$),
and 5 divisions ($/$) in double precision arithmetic,
which sums ($\Sigma$) up to 893 double precision floating-point operations
and averages to 439.3.}
\label{tabopscount}
\end{center}
\end{table}

Parallel algorithms are applied to offset
the cost overhead caused by multiple precision arithmetic.
A specific question asks for the smallest dimension
of the linear system for which teraflop performance is obtained.
Experiencing teraflop performance in quad double arithmetic on a GPU
is similar to about 2.2 gigaflops performance in double arithmetic
on a single threaded execution,
as the average cost of quad double operations is~439.
The 439 is obtained as the average of the $\Sigma$ column
under the quad double header in Table~\ref{tabopscount}.

Double double and quad double arithmetic 
are provided by QDlib~\cite{HLB01},
with its GPU version in~\cite{LHL10}.
The software CAMPARY~\cite{JMPT16} defines
code generators for general multiple float and double 
arithmetical operations.
The handbook~\cite[Chapter~14]{MBDJJLMRT18}
describes multiple double arithmetic.

The Compute to Global Memory Access (CGMA) ratio~\cite{KH10} is
the number of floating-point calculations performed by a kernel
for each access to the global memory.
Looking back at the counts in Table~\ref{tabopscount},
the division of two quad double numbers requires 893
double precision operations on a total of 8 doubles,
naturally leading to a very high CGMA ratio.
An alternative to the CGMA ratio is the roofline model~\cite{WWP09}.
This model is applied in Figure~\ref{figdbl4tabsroofline} to the tiled
accelerated back substitution in quad double precision on the~V100.

The specific motivation for this paper is the development
of a scalable implementation of a new path tracker~\cite{TVV20a}
to solve systems of polynomial equations in several variables.
One component of the path tracker is the solution of a lower
triangular block Toeplitz system~\cite{BV18},
where the diagonal matrix is the evaluated Jacobian matrix
at the current point on the path.
An error analysis in~\cite{TVV20b} motivates the need for 
multiprecision arithmetic if a guaranteed accuracy is desired.
Because of the propagation of roundoff errors, the leading coefficients
in the power series must be computed most accurately,
at a  precision higher than the hardware double precision.
Recently, PHCpack~\cite{Ver99} was extended~\cite{Ver20} with the code
for the multiprecision arithmetic generated by the CAMPARY software,
and applied to accelerate the polynomial evaluation and differentiation
at power series~\cite{Ver21}.

The power series computation provides input to Pad\'{e} approximations,
applied in the holomorphic embedding load flow 
method~\cite{Tri12},~\cite{TM16}, to solve steady state equations
of power systems, using complex analysis.
As indicated in~\cite{RT18}, multiprecision arithmetic
adds significant value. 

\subsection{On Alternatives to CAMPARY}

Compared to genuine multiprecision arithmetic,
multiple double numbers have a limited number of precision levels,
one cannot specify the precise number of bits in the precision.
Another limitation is that the size of the exponents 
are the same as the exponent size of any double.

The authors of~\cite{IK20} compare CAMPARY and CUMP~\cite{NT11}
to their GPU implementation
of multiprecision arithmetic based on the multiple residue number system.
The double double arithmetic of CAMPARY performs best for the problem 
of matrix-vector multiplication.
Concerning quad double precision, the authors of~\cite{IK20} write
``the CAMPARY library is faster than our implementation;
however as the precision increases the execution time of
CAMPARY also increases significantly.''

\subsection{Contributions and Organization}

The main result is the teraflop performance obtained for the multiple
double precision least squares solver, obtained already for relatively
modest dimensions.  
The code generated by the CAMPARY software is applied to solving
linear systems in the least squares sense in double double, quad double,
and octo double precision, for real and complex matrices.
The resulting programs are self contained, available in a github repository,
under the GPL-v3.0 License, thus promoting reproducibility.

The next two sections on accelerating the back substitution
and the blocked Householder QR are meant to provide self-contained
introductions to the parallel implementations
and to explain the legends in the tables in the 
computational experiments section.
Multiple double arithmetic allows for a finer granularity level
as more blocks of threads can collaborate in one matrix-vector product.
The computational experiments start in the fourth section.

\section{Accelerated Back Substitution}

Data parallel algorithms execute the same instructions on different data.
On graphics processing units, this execution is performed by blocks of
threads, scheduled in multiples of~32.
These blocks reside on a number of streaming multiprocessors,
for a total of several thousands of cores.
In order to fully occupy the device, the parallelism must be 
sufficiently fine involving tens of thousands of threads.

In accelerating the back substitution to solve an upper triangular
linear system, the coefficient matrix is divided up into tiles.
The ideas will be illustrated on a 3-by-3 tiled system:

\begin{equation}
   U \x = \bfb, \quad
   U = 
   \left[
      \begin{array}{ccc}
         U_1 & A_{1,2} & A_{1,3} \\
             & U_2 & A_{2,3} \\
             &     & U_3 
      \end{array}
   \right], \quad
   \x = 
   \left[
      \begin{array}{c}
         \x_1 \\ \x_2 \\ \x_3
      \end{array}
   \right], \quad
   \bfb =
   \left[
      \begin{array}{c}
         \bfb_1 \\ \bfb_2 \\ \bfb_3
      \end{array}
   \right],
\end{equation}
where $U_1$, $U_2$, $U_3$ are upper triangular matrices,
with nonzero elements on their diagonal,
and $A_{1,2}$, $A_{1,3}$, and~$A_{2,3}$ are general matrices.
All matrices have the same dimensions.
The length of $\bfb_1$, $\bfb_2$, and $\bfb_3$ equals the
number of rows in each matrix.

In the traditional version of the back substitution algorithm,
the last instruction to compute $x_i$ is the division by the
element on the diagonal.  To introduce more parallelism,
the tiles on the diagonal are first inverted.
The parallel back substitution happens in two stages:
\begin{enumerate}
\item Invert all tiles on the diagonal:
\begin{equation} \label{eqinvertdiag}
   V = 
   \left[
      \begin{array}{ccc}
         U_1^{-1} & A_{1,2} & A_{1,3} \\
             & U_2^{-1} & A_{2,3} \\
             &     & U_3^{-1} 
      \end{array}
   \right].
\end{equation}
The inverse of an upper triangular matrix is again upper triangular.
Each column of the inverse is the solution of an upper triangular system.
The columns of the inverse can be computed independently from each other.
\item The back substitution alternates between multiplying with the inverses
      and updating the right hand side vectors.
The statements on the same line below are executed in parallel.
\begin{eqnarray}
      \x_3 & := & U_3^{-1} \bfb_3, \label{eqbacksubs3} \\
      \bfb_2 & := & \bfb_2 - A_{2,3} \x_3, ~~
      \bfb_1 ~ := ~ \bfb_1 - A_{1,3} \x_3, \\
      \x_2 & := & U_2^{-1} \bfb_2, \label{eqbacksubs2}  \\
      \bfb_1 & := & \bfb_1 - A_{1,2} \x_2, \\
      \x_1 & := & U_1^{-1} \bfb_1.      \label{eqbacksubs1}
\end{eqnarray}
\end{enumerate}
In each step, at least one matrix-vector multiplication is executed.
Each back substitution step requires less work.
With multiple double arithmetic,
the back substitution steps are executed at a finer level: 
multiple blocks of threads cooperate to compute one matrix-vector product.

One could be concerned that the matrix inverse would lead to numerical
instabilities.  However, the tiles are of a much smaller size than the
entire matrix, typically by a factor of at least the number 
of multiprocessors.  Smaller upper triangular matrices have smaller
condition numbers than larger ones.

The example suffices to introduce the main ideas in the algorithm.
To describe the parallelism better,
the accelerated algorithm is presented next in a more formal manner. 

\noindent {\bf Algorithm 1}: {\sc Tiled Accelerated Back Substitution}.

\begin{tabular}{rcl}
  Input & : & $N$ is the number of tiles, \\
        &   & $n$ is the size of each tile, \\
        &   & $U$ is an upper triangular $Nn$-by-$Nn$ matrix, \\
        &   & $\bfb$ is a vector of size $Nn$. \\
 Output & : & $\x$ is a vector of size $Nn$: $U \x = \bfb$.
\end{tabular}
\begin{enumerate}
\item Let $U_1$, $U_2$, $\ldots$, $U_N$ be the $n$-by-$n$ tiles
      on the diagonal of $U$.
Replace each $U_i$ with its inverse $U_i^{-1}$, with $N$ blocks
of $n$ threads.
Labeling threads starting the count at~1,
the $k$-th thread in each block
solves the upper triangular system $U \bfv = \bfe_k$,
where $k$ is the $k$-th $n$-dimensional unit vector.

\item For $i = N, N-1, \ldots, 1$ do

\begin{enumerate}
\item Compute $\x_i := U_i^{-1} \bfb_i$ by one block of $n$ threads.
\item Simultaneously update
      $\bfb_j := \bfb_j - A_{j,i} \x_i$, for $j \in \{ 1,2,\ldots,i-1 \}$,
      with $i-1$ blocks of $n$ threads.
\end{enumerate}

\end{enumerate}
Algorithm~1 executes $1 + N(N+1)/2$ kernel launches.

If $N$ streaming multiprocessors are available and $n$ is a good fit
to keep the device fully occupied, then the computation of all
inverses in the first stage can happen in time proportional to $n^2$,
which is the cost of solving one upper triangular linear system
of dimension~$n$.

Each step in the second stage of Algorithm~1 involves a matrix-vector
multiplication executed in time proportional to $n$,
done by one block of threads.
There are $N$ steps and if sufficiently many multiprocessors are available,
then the total cost of the second stage is proportional to $N n$.
If $n \approx N$, the cost of the first stage can be viewed
as proportional to $N n$, so a good parallel execution of Algorithm~1
can be done in time proportional to $N n$, 
which corresponds to the dimension
of the upper triangular linear system $U \x = \bfb$.

The formulation of Algorithm~1 does not specify the staging of the data.
In particular, the matrix~$U$ of multiple doubles is \underline{\em not}
stored as $U = [u_{i,j}]$, where $u_{i,j}$ is a multiple double,
but as an array $U = [U_1, U_2, \ldots, U_m]$ of $m$ matrices,
where $U_1$ holds the most significant doubles and $U_m$ holds
the least significant doubles.
Similarly, the $\bfb$ in the input of Algorithm~1
is an array of $m$ arrays $[b_1, b_2, \ldots, b_m]$,
ordered in the order of significance.
This facilitates the staggered application of multiprecision 
arithmetic and benefits the efficient memory coalescing:
adjacent threads in one block of
threads read adjacent data in memory, avoiding bank conflicts.
This representation naturally extends to complex arrays,
where the real and imaginary parts are kept separately.

The two questions which will be answered experimentally are
the following.  What is the smallest dimension for which 
teraflop performance is obtained?
Obviously, the lower threshold for $N$ should be the number
of streaming multiprocessors, and $n$ should be a multiple of~32.
The second question asks for the cost overhead factor
in the three times the precision is doubled, from double
to double double, from double double to quad double,
and from quad double to octo double.

\section{Blocked Accelerated Householder QR}

The blocked Householder QR is introduced on a $3m$-by-$3n$ tiled matrix,
$m \geq n$:
\begin{equation}
   A = \left[
     \begin{array}{c|c|c}
              & \multicolumn{2}{c}{A_{1,2}} \\ \cline{2-3}
      A_{1,1} & A_{2,2} & A_{2,3} \\ \cline{3-3}
              &         & A_{3,3}
     \end{array}
   \right], \quad
   \begin{array}{ll}
      A_{1,1} \mbox{ is 3$m$-by-$n$}, \\
      A_{1,2} \mbox{ is $m$-by-2$n$}, \\
      A_{2,2} \mbox{ is 2$m$-by-$n$},  \\
   \end{array}
\end{equation}
$A_{2,3}$ and $A_{3,3}$ are $m$-by-$n$.
The Householder transformations are accumulated in an orthogonal
$3m$-by-$3m$ matrix~$Q$.  The upper triangular reduction~$R$ of~$A$
is written in the matrix~$A$.
The $m$-by-$m$ identity matrix is represented by~$I$
in the sequence of the evolution of $A, Q$ below:
\begin{eqnarray}
  & ~ &
  \left[
     \begin{array}{c|c|c}
              & \multicolumn{2}{c}{A_{1,2}} \\ \cline{2-3}
      A_{1,1} & A_{2,2} & A_{2,3} \\ \cline{3-3}
              &         & A_{3,3}
     \end{array}
  \right], 
  \left[
     \begin{array}{c|c|c}
        I &   & \\
          & I & \\
          &   & I \\
     \end{array}
  \right] \\
  & \!\! \rightarrow \!\! &
  \left[
     \begin{array}{c|c|c}
              & \multicolumn{2}{c}{R_{1,2}} \\ \cline{2-3}
      R_{1,1} & A_{2,2} & A_{2,3} \\ \cline{3-3}
              &         & A_{3,3}
     \end{array}
  \right],
  \left[
     \begin{array}{c|c|c}
          &   & \\
        Q_1  & I & \\
          &   & I \\
     \end{array}
  \right] \\
  & \!\! \rightarrow \!\! &
  \left[
     \begin{array}{c|c|c}
              & \multicolumn{2}{c}{R_{1,2}} \\ \cline{2-3}
      R_{1,1} & R_{2,2} & R_{2,3} \\ \cline{3-3}
              &         & A_{3,3}
     \end{array}
  \right],
  \left[
     \begin{array}{c|c|c}
          &   & \\
        Q_1  & Q_2 & \\
          &   & I \\
     \end{array}
  \right] \\
  & \!\! \rightarrow \!\! &
  \left[
     \begin{array}{c|c|c}
              & \multicolumn{2}{c}{R_{1,2}} \\ \cline{2-3}
      R_{1,1} & R_{2,2} & R_{2,3} \\ \cline{3-3}
              &         & R_{3,3}
     \end{array}
  \right],
  \left[
     \begin{array}{c|c|c}
          &   & \\
        Q_1  & Q_2 & Q_3 \\
          &   &  \\
     \end{array}
  \right].
\end{eqnarray}

One tile $R_{k,k}$ is computed column by column.
For each column, a Householder vector $\bfv$ and corresponding
$\beta = 2/\bfv^T \bfv$ value is computed.
The Householder reflector $P  = I - \beta \bfv \bfv^T$
with the $\bfv$ determined so $P \x = \| \x \|_2 \bfe_1$
(with a sign computation as in \cite[Algorithm 5.1.1]{GV96}),
where $\x$ contains the numbers in the current column
starting at the diagonal, and where $\bfe_1 = (1,0,\ldots,0)^T$.
The Householder matrices are aggregated in an orthogonal matrix
of the form 
\begin{equation}
    P_{WY} = I + W Y^T,
\end{equation}
where $Y$ stores the Householder vectors and has a trapezoidal shape.
The matrix $W$ is computed from the Householder vectors and 
their corresponding $\beta$ values.  With this WY representation
of the Householder matrices, the updates to $Q$ and $R$ can then
written as
\begin{eqnarray}
   Q & = & Q + Q \star W \star Y^T, \\
   R & = & R + Y \star W^T \star C,
\end{eqnarray}
where $C$ is the current matrix to be updated.
The above formulas are rich in matrix-matrix products
which are very suitable for GPU acceleration.
The columns $\z$ of the matrix $W$ follow the formulas
\begin{equation}
   \z = -\beta ( \bfv + W Y^T \bfv),
\end{equation}
which require matrix-vector products.
On complex data, the transpose ${}^T$ is replaced by the
Hermitian transpose ${}^H$.

As stated in~\cite{KCR09}, the computation of $W$ is
expected to be the bottleneck.

The structured description of the accelerated version of the
blocked Householder QR algorithm below serves as an explanation
of the legend of the tables in the next section.

\noindent {\bf Algorithm 2}:
 {\sc Blocked Accelerated Householder QR}.

\begin{tabular}{rcl}
  Input & : & $N$ is the number of tiles, \\
        &   & $n$ is the size of each tile, \\
        &   & $M$ is the number of rows, $M \geq N n$, \\
        &   & $A$ is an $M$-by-$Nn$ matrix. \\
 Output & : & $Q$ is an orthogonal $M$-by-$M$ matrix, \\
        &   & $R$ is an $M$-by-$Nn$ matrix, $A = Q R$.
\end{tabular}

\noindent For $k = 1, 2, \ldots, N$ do
\begin{enumerate}
\item For $\ell = 1, 2, \ldots n$ do 
      \begin{enumerate}
         \item compute $\bfv$ and $\beta$,
         \item update $R_{k,k}$.
      \end{enumerate}
If the size of the current column is less than $n$,
then only one block of threads computes.
Otherwise, several blocks compute $\bfv$, collaborate to update of $R_{k,k}$,
and there is one separate kernel to compute $\beta R^T \star \bfv$,
which also involves a sum reduction with multiple blocks.
\item Given $n$ pairs $(\bfv, \beta)$ computed in the previous stage,
      the matrices $W$, $Y$, and their product $Y \star W^T$ are computed.
\item Update $Q$ in two stages:
\begin{enumerate}
\item $QWY := Q \star WY^T$, where $WY^T = (Y W^T)^T$,
\item $Q := Q + QWY$.
\end{enumerate}
Separating the matrix-matrix multiplication from the addition
clearly shows the cost differences.
In multiple precision arithmetic, the cost of the addition is
however not negligible.
\item If $k < N$, then update $R$, in two stages:
\begin{enumerate}
\item $YWTC := YWT \star C$,
\item $R := R + YWTC$.
\end{enumerate}
\end{enumerate}

As the code is geared towards multiple double arithmetic,
the implementation of the matrix-matrix products differs from 
the double precision implementations recommended in the literature.
When defining kernels for the matrix-matrix multiplication
in double precision, tiles of matrices are loaded into shared memory
to obtain better CGMA ratios, as explained in~\cite[Chapter~5]{KH10}.
Thanks to the high CGMA ratios of multiple double precision,
the entries of the matrix can be loaded directly into the registers
of the kernel that computes one number of the product.

The staging of the data applies the same representation
of multiple double vectors and matrices via multiple arrays of doubles,
as explained at the end of Algorithm~1.

The computational cost of Algorithm~2 is proportional to $M^3$,
if $M = Nn$, for notational simplicity.
If the device is fully occupied, then the hope is to reduce the
cost by a factor of~$M$ and to observe a time proportional to~$M^2$.
In the experiments, as the dimension then doubles, the hope is to
observe the total time multiplied by a factor closer to four
than to eight.

As before, with the back substitution, the first question is to ask
for which dimensions teraflop performance is attained.
The second is to experimentally compute the actual cost overhead factors
of doubling the precision.

\section{Computational Experiments}

In designing the experiments, the first concern is to find sufficiently
large dimensions at which teraflop performance is attained,
mainly in quad double precision.
In the runs at different precisions, 
timings on the double precision version are listed, 
but are not used in the comparisons as the implementation
was made for multiple double precision arithmetic.
Another reason for not comparing the timings of runs in double precision
is that the dimensions are not yet large enough to fully occupy the device.

In a first comparison of runs at different precisions, 
the tile size and the corresponding number of threads per block 
are fixed to the same number for all precisions.  
However, in double double precision, the number of threads per block 
should be higher than in octo double precision.

In all tables, all time units are milliseconds.  
The units of flops (floating-point operations per second) are gigaflops.

\subsection{Notes on the Implementation}

The code for the multiple double arithmetical operations
generated by the CAMPARY software~\cite{JMPT16}
was customized for each precision in the following manner.
Instead of representing a quad double number by an array of four doubles,
all arithmetical operations work on four separate variables,
one for each double.
By this customization an array of quad doubles is stored as
four separate arrays of doubles and a matrix of quad doubles 
is represented by four matrices of doubles.
If one would be only interested in double double and quad double,
then the {\tt double2} and {\tt double4} types of the CUDA SDK
will work just as well (as we did in~\cite{VY13}),
but then performance drops are to be
expected with complex quad doubles already and then also 
for the more general multiple double arithmetic.

QDlib~\cite{HLB01}
provides definitions for the square roots and various other
useful functions for double double and quad double arithmetic.
Those definitions are extended to octo double precision,
also with the customization of representing an octo double number
as eight different variables.

The {\tt \_\_forceinline\_\_} directive was added to all the
device functions that define the multiple double arithmetic.
All {\tt .cu} files are compiled with {\tt nvcc -O3}.

For every kernel in the implementation of Algorithms~1 and~2,
a small function accumulates the number of arithmetical operations.
Then the total number of floating-point operations is computed
at the end of the run,
using the numbers in Table~\ref{tabopscount} as multipliers.
In the application of the roofline model, 
the number of bytes in each computation is obtained 
from the dimensions of the problem,
multiplied by the size of each multiple double number.

Random numbers were generated for the input matrices.
In the standalone tests on the back substitution solver,
the random upper triangular matrices were computed on the host 
as the output of an LU factorization,
as the condition numbers of random triangular matrices almost surely
grow exponentially~\cite{VT98}.
All tests were run on well conditioned problems,
so the residuals $\| \bfb - A \x \|_2^2$ of the computed solution~$\x$
to the linear system $A \x = \bfb$ is of the expected accuracy,
corresponding to the level of the multiple double precision.

The same code runs on five different NVIDIA GPUs.
The C2050, K20C, P100, V100 are housed in CentOS workstations
and the gcc compiler is used to compile the code on the host.
The RTX 2080 resides in a Windows laptop, and
the community edition of Microsoft Visual Studio is used.

The code is free and open source, released under the GPU GPL license,
in the PHCpack source available on 
{\small \tt https://github.com/janverschelde/PHCpack}.

\subsection{Equipment}

Using the same setup as in~\cite{Ver21},
Table~\ref{tabgpus} lists the main characteristics of five GPUs
used to develop the code.

\begin{table}[hbt]
\begin{center}
\begin{tabular}{r||r|r|r|r|r||r}
  NVIDIA GPU~~~ &  CUDA  & \#MP  & \#cores/MP & \#cores & GHz 
 & host CPU GHz~~ \\ \hline
      Tesla C2050 & 2.0~~ &  14~~ &    32~~~~~ &   448~ & 1.15 
 & Intel X5690 3.47 \\
      Kepler K20C & 3.5~~ &  13~~ &   192~~~~~ &  2496~ & 0.71 
 & Intel E5-2670 2.60 \\
      Pascal P100 & 6.0~~ &  56~~ &    64~~~~~ &  3584~ & 1.33 
 & Intel E5-2699 2.20 \\
      Volta V100 & 7.0~~ &  80~~ &    64~~~~~ &  5120~ & 1.91 
 & Intel W2123 3.60 \\
 GeForce RTX 2080 & 7.5~~ &  46~~ &    64~~~~~ &  2944~ & 1.10 
 & Intel i9-9880H 2.30
\end{tabular}
\caption{The columns list the CUDA capability, 
the number of multiprocessors, the number of cores per multiprocessor,
the total number of cores, and the GPU clock rate.
For every GPU, its host CPU is listed with its clock rate,
and the host processor.}
\label{tabgpus}
\end{center}
\end{table}

While running the same software on different GPUs is convenient,
the obvious disadvantage is that the more advanced features of
the newer devices are not utilized.
Important in the investigation of the scalability is the
attention to teraflop performance and the ratios of the theoretical
peak performances of the V100 over the P100.

In each run, 
the elapsed times of the kernel launches are measured by
{\tt cudaEventElapsedTime} and are expressed in milliseconds.
The wall clock times include the sum of times spent by the kernels,
with the added memory transfers.
The kernel flops in the tables below are the totals of
the counts of the double precision operations over the sum of
the times spent by the kernels. 
The total wall clock time is used in the wall flops.

\subsection{Blocked Householder QR on Five Different GPUs}

The theoretical double peak performance of the P100 and the V100
are 4.7 TFLOPS and 7.9 TFLOPS respectively.
Therefore, if the code scales well,
one may expect the V100 to be about 1.68 times faster than the P100.

\begin{table}[t!]
\begin{center}
\begin{tabular}{r||r|r|r|r||r}
   stage in       & \multicolumn{4}{c||}{Linux on the host}
                  & \multicolumn{1}{c}{Windows} \\
   Algorithm~2 
  & C2050  &  K20C  &  P100  &  V100  & RTX 2080 \\ \hline \hline
 $\beta, v$         &   35.5 &   43.8 &   21.4 &   16.2 &   26.2~ \\
$\beta R^T \star v$ &  418.8 &  897.8 &   89.6 &   76.6 &  389.7~ \\
 update $R$         &  107.0 &  107.6 &   23.0 &   15.2 &   47.5~ \\
 compute $W$        & 1357.8 & 1631.8 &  349.2 &  222.4 & 1298.4~ \\
 $Y \star W^T$      &  100.0 &   50.3 &    9.7 &    6.6 &  153.5~ \\
 $Q \star WY^T$     &  790.9 &  423.9 &   77.2 &   52.1 & 1228.8~ \\
 $YWT \star C$      & 6068.5 & 2345.2 &  141.2 &   61.6 &  822.6~ \\
 $Q + QWY$          &    2.4 &    1.6 &    0.4 &    0.4 &    0.7~ \\
 $R + YWTC$         &    7.4 &    4.2 &    0.7 &    0.5 &    0.8~ \\ \hline
 all kernels        & 8888.3 & 5506.1 &  712.4 &  451.5 & 3968.2~ \\
 wall clock         & 9083.0 & 5682.0 &  826.0 &  568.0 & 4700.0~ \\ \hline
kernel flops        & 115.8  &  187.0 & 1445.3 & 2280.4 &  259.5~ \\
 wall flops         & 113.4  &  181.2 & 1247.2 & 1812.7 &  219.1~
\end{tabular}
\caption{Blocked Householder QR in double double precision,
on a 1,024-by-1,024 matrix, with 8 tiles of size 128.}
\label{tab5gpus}
\end{center}
\end{table}

For the total kernel time in Table~\ref{tab5gpus},
compare the scaled observed time on the V100: 
$451.5 \times 1.68 \approx 758.5$,
with the observed 712.4 of the P100.
Comparing wall clock times is harder,
because of different clock speeds of the host processor
and the workstation that hosts the P100 has 256GB of RAM,
whereas the RAM of the host of the V100 holds~32GB.

For historical perspective, the oldest C2050 was purchased in 2011
and the V100 in 2019.  The ratio of the sum of the times
spent on all kernels of the C2050 over V100: $8888.3/451.5 \approx 19.6$,
indicating about a double speedup every two years.
The tile size of 128 (and the number of threads per block)
is most likely not the best choice for the K20C,
which has 192 cores per streaming multiprocessor.
We used the K20C in~\cite{VY13}.
In the single experiment comparison in Table~\ref{tab5gpus},
the GeForce RTX 2080 Max-Q outperforms the K20C.

\subsection{Blocked Householder QR in Four Different Precisions}

Based on the operational counts in Table~\ref{tabopscount},
one could predict the overhead factors from the averages
in the $\Sigma$ column, which are 37.7 for double double,
439.3 for quad double, and 2379.0 for octo double.
Based on those averages, going from double double to quad double
would cause all times to be multiplied by 11.7.
Similarly, the predicted overhead factor is 5.4
when going from quad double to octo double.


\begin{figure}[hbt]
\begin{center}
\includegraphics[width=9cm]{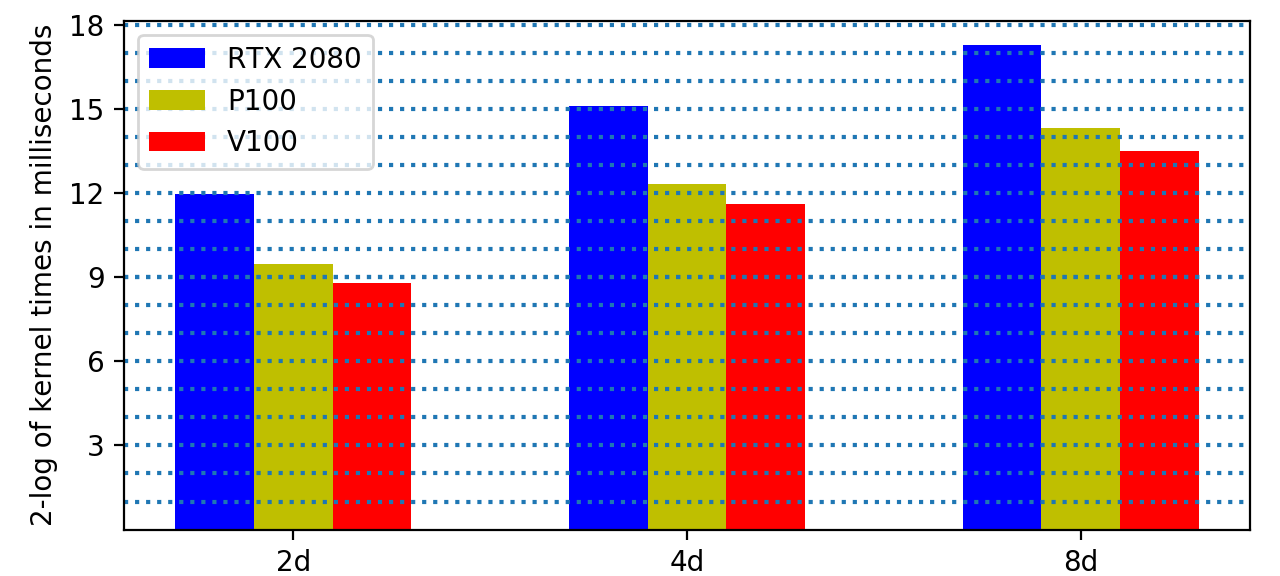}
\caption{2-logarithms of the times spent by all kernels of QR
on the RTX~2080, the P100, and the V100 in double double (2d),
quad double (4d), and octo double (8d) precision,
with data in Table~\ref{tabqr4prc}.}
\label{figqr4prc}
\end{center}
\end{figure}

Table~\ref{tabqr4prc} illustrates the cost overhead
of doubling the precision three times and is summarized
by Figure~\ref{figqr4prc}.
From double double to quad double,
taking ratios of kernels times $5187.0/712.7 \approx 7.3$ on the P100, and
the similar ratio on the V100 is $3167.0/446.8 \approx 7.1$.
Both ratios are consistent and less than the predicted factor of~11.7.
On the RTX 2080, the ratio is $35826.7/3999.5 \approx 9.0 < 11.7$.
From quad double to octo double,
the kernels time ratio on the P100 is $20547.5/5187.0 \approx 4.0$
and $11754.6/3167.0 \approx 3.7$.  In both cases, the observed factors
are less than the predicted~5.4,
as is also the case on the RTX~2080: $160802.8/35826.7 \approx 4.5$.
That the observed cost overhead factors are more favorable than
the predicted ones correlates with the increased performance
for increased precisions.

\begin{table}[t!]
\begin{center}
{\small
\begin{tabular}{r||r|r|r|r}
 \multicolumn{1}{c||}{stage in}
  & \multicolumn{4}{c}{times on the RTX 2080} \\
 Algorithm~2 
    & \multicolumn{1}{c|}{1d} 
    & \multicolumn{1}{c|}{2d}
    & \multicolumn{1}{c|}{4d}
    & \multicolumn{1}{c}{8d} \\ \hline \hline
 $\beta, v$         &  13.0 &   26.3 &   108.1 &    451.8 \\
$\beta R^T \star v$ &  46.3 &  338.0 &  1740.9 &   4994.3 \\
  update $R$        &  11.3 &   47.7 &   376.9 &   1669.6 \\
  compute $W$       & 111.7 & 1309.4 & 12346.8 &  56484.2 \\
 $Y \star W^T$      &   5.1 &  154.7 &  1476.3 &   6746.7 \\
 $Q \star WY^T$     &  40.2 & 1238.7 & 11815.5 &  54008.9 \\
 $YWT \star C$      & 110.1 &  833.3 &  7957.3 &  36430.5 \\
 $Q + QWY$          &   0.3 &    0.7 &     3.1 &      9.4 \\
 $R + YWTC$         &   0.5 &    0.8 &     1.9 &      7.3 \\ \hline
 all kernels        & 338.6 & 3999.5 & 35826.7 & 160802.8 \\
 wall clock         & 562.0 & 4708.0 & 37087.0 & 163219.0 \\ \hline
kernel flops        & 141.5 &  257.4 &   284.1 &    299.7 \\
 wall flops         &  85.2 &  218.7 &   274.5 &    295.3 \\ \hline \hline
 \multicolumn{1}{c||}{stage in}
  & \multicolumn{4}{c}{times on the P100} \\
 Algorithm~2 
    & \multicolumn{1}{c|}{1d} 
    & \multicolumn{1}{c|}{2d}
    & \multicolumn{1}{c|}{4d}
    & \multicolumn{1}{c}{8d} \\ \hline \hline
 $\beta, v$         &  12.6 &   21.6 &   58.3 &   412.4 \\
$\beta R^T \star v$ &  44.4 &   89.7 &  760.7 &  2998.5 \\
  update $R$        &  14.2 &   23.0 &   96.3 &   359.9 \\
  compute $W$       &  98.3 &  349.3 & 2752.3 &  9857.5 \\
 $Y \star W^T$      &   3.0 &    9.7 &   96.8 &   484.7 \\
 $Q \star WY^T$     &  25.0 &   77.0 &  747.0 &  3745.4 \\
 $YWT \star C$      &  67.1 &  141.3 &  672.8 &  2681.7 \\
 $Q + QWY$          &   0.2 &    0.4 &    0.9 &     2.0 \\
 $R + YWTC$         &   0.4 &    0.7 &    1.8 &     5.5 \\ \hline
 all kernels        & 256.2 &  712.7 & 5187.0 & 20547.5 \\
 wall clock         & 311.0 &  827.0 & 5381.0 & 20870.0 \\ \hline
kernel flops        & 180.6 & 1444.6 & 1962.4 &  2345.4 \\
 wall flops         & 154.0 & 1244.8 & 1891.5 &  2309.2 \\ \hline \hline
 \multicolumn{1}{c||}{stage in}
  & \multicolumn{4}{c}{times on the V100} \\
 Algorithm~2
    & \multicolumn{1}{c|}{1d} 
    & \multicolumn{1}{c|}{2d}
    & \multicolumn{1}{c|}{4d}
    & \multicolumn{1}{c}{8d} \\ \hline \hline
 $\beta, v$         &   7.3 &   15.8 &   37.4 &   180.2 \\
$\beta R^T \star v$ &  28.8 &   77.2 &  470.4 &  1304.0 \\
  update $R$        &   9.4 &   15.1 &   58.9 &   197.9 \\
  compute $W$       &  79.5 &  223.2 & 1551.0 &  5700.9 \\
 $Y \star W^T$      &   0.8 &    6.5 &   66.3 &   281.3 \\
 $Q \star WY^T$     &   8.2 &   56.7 &  516.8 &  2249.2 \\
 $YWT \star C$      &  24.0 &   51.4 &  464.4 &  1834.5 \\
 $Q + QWY$          &   0.2 &    0.4 &    0.8 &     1.6 \\
 $R + YWTC$         &   0.2 &    0.4 &    1.0 &     4.8 \\ \hline
 all kernels        & 158.4 &  446.8 & 3167.0 & 11754.6 \\
 wall clock         & 206.0 &  560.0 & 3356.0 & 12059.0 \\ \hline
kernel flops        & 302.5 & 2304.3 & 3214.0 &  4099.9 \\
 wall flops         & 232.8 & 1837.3 & 3033.0 &  3996.3
\end{tabular}
}
\caption{Blocked Householder QR in double (1d), double double (2d),
quad double (4d), and octo double (8d) precision,
on a 1,024-by-1,024 matrix, with 8 tiles of size 128,
on the RTX~2080, the P100, and the V100.}
\label{tabqr4prc}
\end{center}
\end{table}

\subsection{Real and Complex Double Double QR}

Working with complex arithmetic requires about four times
as many arithmetical operations than on real data.
Table~\ref{tabV100RCdd} lists times on 512-by-512 matrices,
of real and complex double double numbers.
Keeping the dimension 512 constant, 
fewer tiles but larger tiles are selected with each execution.

\begin{table}[t!]
\begin{center}
\begin{tabular}{r||r|r|r|r}
 \multicolumn{1}{c||}{stage in}
  & \multicolumn{4}{c}{on real matrices} \\
 Algorithm~2 
    & \multicolumn{1}{c|}{$16 \times 32$} 
    & \multicolumn{1}{c|}{$8 \times 64$}
    & \multicolumn{1}{c|}{$4 \times 128$}
    & \multicolumn{1}{c}{$2 \times 256$} \\ \hline \hline
 $\beta, v$         &   6.5 &  10.7 &    7.8 &   7.7 \\
$\beta R^T \star v$ &  12.4 &  22.0 &   20.2 &  20.0 \\
  update $R$        &   2.3 &   4.9 &    9.9 &  46.6 \\
  compute $W$       &  22.9 &  41.9 &   54.1 &  81.7 \\
 $Y \star W^T$      &   0.5 &   0.9 &    1.0 &   1.1 \\
 $Q \star WY^T$     &   4.3 &   7.0 &    3.9 &   2.8 \\
 $YWT \star C$      &   4.0 &   6.3 &    3.6 &   1.5 \\
 $Q + QWY$          &   0.1 &   0.1 &    0.1 &   0.1 \\
 $R + YWTC$         &   0.1 &   0.1 &    0.1 &   0.1 \\ \hline
 all kernels        &  53.2 &  94.0 &  100.5 & 161.6 \\
 wall clock         & 101.0 & 170.0 &  155.0 & 208.0 \\ \hline
kernel flops        & 428.4 & 785.9 & 1089.8 & 777.3 \\
 wall flops         & 226.6 & 434.5 &  707.4 & 603.3 \\ \hline \hline
 \multicolumn{1}{c||}{stage in}
  & \multicolumn{4}{c}{on complex matrices} \\
 Algorithm~2 
    & \multicolumn{1}{c|}{$16 \times 32$} 
    & \multicolumn{1}{c|}{$8 \times 64$}
    & \multicolumn{1}{c|}{$4 \times 128$}
    & \multicolumn{1}{c}{$2 \times 256$} \\ \hline \hline
 $\beta, v$         &   8.5 &    8.4 &    8.3 &    8.9 \\
$\beta R^T \star v$ &  20.6 &   36.8 &   36.7 &   37.3 \\
  update $R$        &   3.0 &    6.8 &   20.5 &  204.7 \\
  compute $W$       &  38.9 &  126.6 &  144.3 &  248.9 \\
 $Y \star W^T$      &   0.9 &    3.3 &    3.7 &    4.5 \\
 $Q \star WY^T$     &  12.7 &   26.4 &   15.1 &   11.3 \\
 $YWT \star C$      &  12.4 &   18.6 &    9.8 &    5.1 \\
 $Q + QWY$          &   0.2 &    0.2 &    0.1 &    0.1 \\
 $R + YWTC$         &   0.3 &    0.2 &    0.1 &    0.1 \\ \hline
 all kernels        &  97.4 &  227.4 &  238.5 &  420.8 \\
 wall clock         & 158.0 &  306.0 &  311.0 &  479.0 \\ \hline
kernel flops        & 628.9 & 1299.8 & 1836.7 & 1194.8 \\
 wall flops         & 387.2 &  967.3 & 1407.8 & 1050.5 \\
\end{tabular}
\caption{Blocked Householder QR in double double precision,
on real and complex matrices of dimension 512,
for increasing tile sizes,
$512 = 16 \times 32 = 8 \times 64 = 4 \times 128 = 2 \times 256$, 
on the V100.}
\label{tabV100RCdd}
\end{center}
\end{table}

Looking at the flops in Table~\ref{tabV100RCdd},
teraflop performance is reached for both real and complex matrices,
for 128 as the tile size.  
At $4 \times 128$, the device is best occupied.
But if interested in total wall clock times,
then $16 \times 32$ is best.

In a dimension as small as 512, the computation of $W$ dominates.
Would this still be the case if the dimensions increase?

\subsection{Quad Double QR for Increasing Dimensions}

\begin{figure}[hbt]
\begin{center}
{\includegraphics[width=9cm]{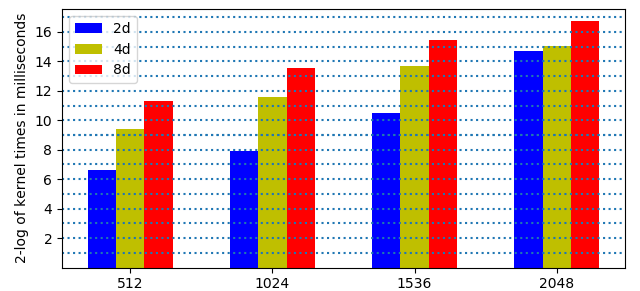}}
\caption{2-logarithms of the times spent by all kernels of QR
on the V100 in double double (2d), quad double (4d), 
and octo double (8d) precision, for increasing dimensions,
for the data in Table~\ref{tabV100baqr4d}.}
\label{figV100baqrprc}
\end{center}
\end{figure}

How do the execution times of the QR decomposition
evolve for increasing dimensions?
Table~\ref{tabV100baqr4d} lists the times 
for dimensions 512, 1024, 1536, and 2048.
Figure~\ref{figV100baqrprc} shows the evolution of all kernel times.

\begin{table}[t!]
\begin{center}
\begin{tabular}{r||r|r|r|r}
  & \multicolumn{4}{c}{double double precision} \\
 \multicolumn{1}{c||}{stage in}
  & \multicolumn{1}{c|}{512} 
  & \multicolumn{1}{c|}{1024}
  & \multicolumn{1}{c|}{1536}
  & \multicolumn{1}{c}{2048} \\
 Algorithm~2 
    & \multicolumn{1}{c|}{$4 \times 128$} 
    & \multicolumn{1}{c|}{$8 \times 128$}
    & \multicolumn{1}{c|}{$12 \times 128$}
    & \multicolumn{1}{c}{$16 \times 128$} \\ \hline \hline
 $\beta, v$         &    7.9 &    8.2 &   16.5 &    34.5 \\
$\beta R^T \star v$ &   20.3 &   36.7 &  144.6 &   652.0 \\
  update $R$        &    9.9 &   20.5 &   28.9 &    58.8 \\
  compute $W$       &   53.9 &  144.2 &  556.2 &  2278.7 \\
 $Y \star W^T$      &    1.0 &    3.7 &   24.9 &   194.2 \\
 $Q \star WY^T$     &    4.0 &   15.1 &  201.3 &  3048.9 \\
 $YWT \star C$      &    3.4 &    9.7 &  481.6 & 20534.4 \\
 $Q + QWY$          &    0.1 &    0.1 &    0.7 &     5.6 \\
 $R + YWTC$         &    0.1 &    0.1 &    0.9 &     7.8 \\ \hline
 all kernels        &  100.5 &  238.2 & 1455.8 & 26815.0 \\
 wall clock         &  155.0 &  321.0 & 1627.0 & 27230.0 \\ \hline
kernel flops        & 1089.7 & 1839.0 & 2475.1 &  1087.8 \\
 wall flops         &  706.5 & 1364.9 & 2214.4 &  1071.2 \\ \hline \hline
 & \multicolumn{4}{c}{quad double precision} \\
 \multicolumn{1}{c||}{stage in}
  & \multicolumn{1}{c|}{512} 
  & \multicolumn{1}{c|}{1024}
  & \multicolumn{1}{c|}{1536}
  & \multicolumn{1}{c}{2048} \\
 Algorithm~2 
    & \multicolumn{1}{c|}{$4 \times 128$} 
    & \multicolumn{1}{c|}{$8 \times 128$}
    & \multicolumn{1}{c|}{$12 \times 128$}
    & \multicolumn{1}{c}{$16 \times 128$} \\ \hline \hline
 $\beta, v$         &   21.0 &   37.4 &    54.1 &    71.6 \\
$\beta R^T \star v$ &  115.5 &  470.9 &  1073.8 &  1939.9 \\
  update $R$        &   49.0 &   59.0 &    74.6 &    91.2 \\
  compute $W$       &  412.6 & 1553.5 &  3438.2 &  6104.3 \\
 $Y \star W^T$      &    9.6 &   66.3 &   214.2 &   517.6 \\
 $Q \star WY^T$     &   41.5 &  538.3 &  2511.0 &  7643.9 \\
 $YWT \star C$      &   24.9 &  409.3 &  6057.1 & 17991.2 \\
 $Q + QWY$          &    0.1 &    0.8 &     2.5 &     5.7 \\
 $R + YWTC$         &    0.1 &    0.9 &     5.6 &     7.0 \\ \hline
 all kernels        &  674.3 & 3136.5 & 13431.2 & 34372.5 \\
 wall clock         &  777.0 & 3366.0 & 13835.0 & 34960.0 \\ \hline
kernel flops        & 1605.7 & 3245.3 &  2366.8 &  2097.0 \\
 wall flops         & 1392.6 & 3024.4 &  2297.7 &  2061.7 \\ \hline \hline
 & \multicolumn{4}{c}{octo double precision} \\
 \multicolumn{1}{c||}{stage in}
  & \multicolumn{1}{c|}{512} 
  & \multicolumn{1}{c|}{1024}
  & \multicolumn{1}{c|}{1536}
  & \multicolumn{1}{c}{2048} \\
 Algorithm~2 
    & \multicolumn{1}{c|}{$4 \times 128$} 
    & \multicolumn{1}{c|}{$8 \times 128$}
    & \multicolumn{1}{c|}{$12 \times 128$}
    & \multicolumn{1}{c}{$16 \times 128$} \\ \hline \hline
 $\beta, v$         &   94.7 &   188.3 &   282.8 &    385.1 \\
$\beta R^T \star v$ &  309.5 &  1309.1 &  3009.7 &   5416.5 \\
  update $R$        &  167.5 &   199.0 &   245.1 &    300.4 \\
  compute $W$       & 1568.2 &  5828.6 & 12908.6 &  22944.3 \\
 $Y \star W^T$      &   48.3 &   308.8 &   957.5 &   2082.8 \\
 $Q \star WY^T$     &  187.2 &  2334.8 & 11259.2 &  34508.6 \\
 $YWT \star C$      &  114.9 &  2104.0 & 15982.0 &  42044.8 \\
 $Q + QWY$          &    0.3 &     1.6 &     5.1 &     11.6 \\
 $R + YWTC$         &    0.2 &     5.9 &    29.8 &     75.2 \\ \hline
 all kernels        & 2490.8 & 12280.1 & 44679.8 & 107769.2 \\
 wall clock         & 2681.0 & 12735.0 & 45419.0 & 108763.0 \\ \hline
kernel flops        & 2058.2 &  3924.4 &  3368.5 &   3166.4 \\
 wall flops         & 1912.0 &  3784.2 &  3313.6 &   3137.5 \\
\end{tabular}
\caption{Blocked Householder QR 
in double double, quad double, and octo double precision,
on real matrices of increasing dimensions,
for increasing number of tiles,
$512 = 4 \times 128$, $1024 = 8 \times 128$,
$1536 = 12 \times 128$, $2048 = 16 \times 128$, on the V100.}
\label{tabV100baqr4d}
\end{center}
\end{table}

At dimension~512, the computation of~$W$ dominates in all precisions.
The accumulated times of all kernels devoted to computing~$W$
drops to the thirdmost largest time in dimension~2048.
The two most time consuming kernels are those that do
the matrix-matrix multiplications.

Doubling the dimension, from 512 to 1024, the ratios of the wall
clock times in double double, quad double, and octo double precision
are respectively $321.0/155.0 \approx 2.1$,
$3366.0/777.0 \approx 4.3$, and $12735.0/2681.0 \approx 4.8$,
corresponding to significant increases in performance.

While teraflop performance is maintained, notice
in Table~\ref{tabV100baqr4d} the drop in performance
at dimension~2048 in double double arithmetic.
This drop is most likely due to the kernels for the matrix-matrix
multiplication that do not take advantage of the shared memory,
as the double double arithmetic has not yet a high enough CGMA ratio,
compared to the higher multiple double arithmetic.
Although not as much as in double double precision,
there is also a drop in performance in the other precisions 
for the two largest dimensions.

\subsection{Back Substitution in Four Different Precisions}

The high CGMA ratios makes that the overhead cost of doubling 
the precisions is less than the predicted overhead factors.
Would this also be the case for the back substitution?

Consider the doubling of 
both the dimension and the precision.
Table~\ref{tabbs4prc} records the times
of the back substitution on upper triangular matrices of sizes
$5120 = 64 \times 80$, $10240 = 128 \times 80$, and $20480 = 256 \times 80$, 
where the first factors in the dimensions are the size of each tile
and the second factors are the number of tiles.
In octo double precision, shared memory capacities limit the tile size
to 128, so then $20480 = 128 \times 160$.
The high wall clock time in octo double precision for 20480 is due
to the limited 32 GB of RAM at the host.
Despite this anomaly, the times spent by all kernels appear regular
enough to reliably measure the cost overhead factors from doubling
the precisions.

\begin{table}[htb]
\begin{center}
\begin{tabular}{r||r|r|r}
\multicolumn{4}{c}{double precision} \\
  stage in Algorithm~1 
 & $64 \times 80$ & $128 \times 80$ & $256 \times 80$ \\ \hline \hline
 invert diagonal tiles &   0.4~~ &   5.2~~ &  30.8~~ \\
multiply with inverses &   0.8~~ &   1.5~~ &   4.3~~ \\
     back substitution &   1.8~~ &   2.2~~ &   5.9~~ \\ \hline
 time spent by kernels &   3.0~~ &   8.9~~ &  41.0~~ \\ 
       wall clock time &  47.0~~ & 147.0~~ & 526.0~~ \\ \hline \hline
     kernel time flops &  14.5~~ &  28.5~~ &  39.9~~ \\
      wall clock flops &   0.9~~ &   1.7~~ &   3.1~~ \\ \hline \hline
\multicolumn{4}{c}{double double precision} \\
  stage in Algorithm~1 
 & $64 \times 80$ & $128 \times 80$ & $256 \times 80$ \\ \hline \hline
 invert diagonal tiles &   1.2~~ &   9.3~~ &   46.3~~ \\
multiply with inverses &   1.7~~ &   3.3~~ &    8.9~~ \\
     back substitution &   7.9~~ &   4.7~~ &   12.2~~ \\ \hline
 time spent by kernels &   5.0~~ &  17.3~~ &   67.4~~ \\
       wall clock time &  82.0~~ & 286.0~~ &  966.0~~ \\ \hline \hline
     kernel time flops & 190.6~~ & 318.7~~ &  525.1~~ \\
      wall clock flops &  11.7~~ &  19.2~~ &   36.7~~ \\ \hline \hline
\multicolumn{4}{c}{quad double precision} \\
  stage in Algorithm~1 
 & $64 \times 80$ & $128 \times 80$ & $256 \times 80$ \\ \hline \hline
 invert diagonal tiles &   6.2~~ &  38.3~~ &  137.4~~ \\
multiply with inverses &  12.2~~ &  23.8~~ &   63.1~~ \\
     back substitution &  13.3~~ &  26.7~~ &  112.2~~ \\ \hline
 time spent by kernels &  31.7~~ &  88.8~~ &  312.7~~ \\
       wall clock time & 187.0~~ & 619.0~~ & 2268.0~~ \\ \hline \hline
     kernel time flops & 299.4~~ & 614.2~~ & 1122.3~~ \\
      wall clock flops &  50.8~~ &  88.1~~ &  154.8~~ \\ \hline \hline
\multicolumn{4}{c}{octo double precision} \\
  stage in Algorithm~1 
 & $64 \times 80$ & $128 \times 80$ & $128 \times 160$ \\ \hline \hline
 invert diagonal tiles &  43.8~~ &  110.6~~ &   133.3~~ \\
multiply with inverses &  47.7~~ &   97.5~~ &   196.0~~ \\
     back substitution &  49.2~~ &  108.0~~ &   283.7~~ \\ \hline
 time spent by kernels & 140.7~~ &  316.2~~ &   613.1~~ \\
       wall clock time & 465.0~~ & 1400.0~~ & 84448.0~~ \\ \hline \hline
     kernel time flops & 321.3~~ &  820.1~~ &  1166.7~~ \\
      wall clock flops &  97.1~~ &  185.2~~ &     8.5~~
\end{tabular}
\caption{Back substitution in four different precisions on
problems of increasing size, on the V100.}
\label{tabbs4prc}
\end{center}
\end{table}

In double precision, the times spent by the kernels are not large
enough to attain a good performance.
At the largest dimension,
half a teraflop is reached in double double precision;
in quad and octo double precision, 1.1 teraflop is observed.

\begin{figure}[hbt]
\begin{center}
{\includegraphics[width=9cm]{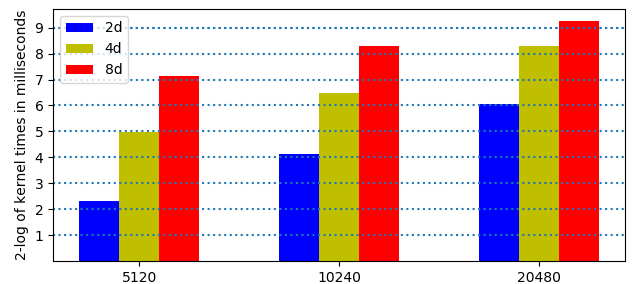}}
\caption{2-logarithms of the times in Table~\ref{tabbs4prc}
spent by all kernels of back substitution on the V100, 
for dimension 5120, 10240, 20480,
in double (1d), double double (2d),
quad double (4d), and octo double (8d) precision.}
\label{figbs4prc}
\end{center}
\end{figure}

Figure~\ref{figbs4prc} shows the 2-logarithms of the times
spent by all kernels.
As the cost of the back substitution is quadratic in the dimension,
the times are expected to quadruple when the dimension is doubled.
This quadrupling is observed in double double precision,
but then becomes closer to doubling in octo double precision, 
thanks to the higher performance in higher precision.
Observe that the heights of the quad double bar is closer 
to the octo double bar than to the double double bar.  
This is consistent with the predicted cost overhead ratios, 
which are higher when going from double double to quad double
compared to the ratios from quad double to octo double.

\begin{table}[htb]
\begin{center}
\begin{tabular}{r||r|r|r}
\multicolumn{1}{c||}{stage in Algorithm~1}
 & \multicolumn{1}{c|}{$320 \times 64$} 
 & \multicolumn{1}{c|}{$160 \times 128$}
 & \multicolumn{1}{c}{$80 \times 256$} \\ \hline \hline
 invert diagonal tiles &   13.5~~ &   35.8~~ &  132.3~~ \\
multiply with inverses &   49.0~~ &   47.5~~ &   64.3~~ \\
     back substitution &   84.6~~ &   91.7~~ &  112.3~~ \\ \hline
 time spent by kernels &  147.1~~ &  175.0~~ &  308.9~~ \\
       wall clock time & 2620.0~~ & 2265.0~~ & 2071.0~~ \\ \hline \hline
     kernel time flops &  683.0~~ &  861.1~~ & 1136.1~~ \\
      wall clock flops &   38.3~~ &   66.5~~ &  169.5~~
\end{tabular}
\caption{Back substitution in quad double precision
in dimension~$20480 = N \times n$,
for three different combinations of $N$ and $n$, on the V100. }
\label{tabV100nofixNtabs4}
\end{center}
\end{table}

\begin{table}[htb]
\begin{center}
\begin{tabular}{r||r|r|r|r|r|r|r|r}
& \multicolumn{8}{c}{times on the RTX 2080} \\
  \multicolumn{1}{c||}{stage in Algorithm~1}
  & \multicolumn{1}{c|}{$32$} 
  & \multicolumn{1}{c|}{$64$} 
  & \multicolumn{1}{c|}{$96$} 
  & \multicolumn{1}{c|}{$128$}
  & \multicolumn{1}{c|}{$160$} 
  & \multicolumn{1}{c|}{$192$}
  & \multicolumn{1}{c|}{$224$}
  & \multicolumn{1}{c}{$256$} \\ \hline \hline
  invert diagonal tiles &  14.7 & 101.1 & 272.0 &  460.0 &  762.8 & 1163.6 & 1758.5 & 1589.3 \\
 multiply with inverses &  41.7 &  67.0 & 104.5 &  184.3 &  293.2 &  416.1 &  556.4 &  747.6 \\
      back substitution &  50.4 &  99.6 & 147.9 &  263.0 &  409.1 &  590.7 &  781.4 & 1055.4 \\ \hline
  time spent by kernels & 106.8 & 267.7 & 524.4 &  907.2 & 1465.1 & 2170.4 & 3096.3 & 4392.3 \\
        wall clock time & 174.0 & 420.0 & 883.0 & 1477.0 & 2318.0 & 3343.0 & 4725.0 & 6726.0 \\ \hline \hline
      kernel time flops &  17.4 &  35.5 &  49.6 &   60.1 &   67.0 &   73.8 &   78.6 &   79.9 \\
       wall clock flops &  10.7 &  22.6 &  29.5 &   37.0 &   42.4 &   47.9 &   51.5 &   52.2 \\ \hline \hline
 & \multicolumn{8}{c}{times on the P100} \\
 \multicolumn{1}{c||}{stage in Algorithm~1}
  & \multicolumn{1}{c|}{$32$} 
  & \multicolumn{1}{c|}{$64$} 
  & \multicolumn{1}{c|}{$96$} 
  & \multicolumn{1}{c|}{$128$}
  & \multicolumn{1}{c|}{$160$} 
  & \multicolumn{1}{c|}{$192$}
  & \multicolumn{1}{c|}{$224$}
  & \multicolumn{1}{c}{$256$} \\ \hline \hline
  invert diagonal tiles &   2.3 &   8.6 &  18.9 &   35.6 &   61.0 &   97.2 &  148.1 &  215.6 \\
 multiply with inverses &  11.0 &  20.4 &  29.5 &   40.0 &   51.0 &   64.3 &   74.3 &   89.3 \\
      back substitution &  10.9 &  20.6 &  30.3 &   43.5 &   64.4 &   98.3 &  109.8 &  126.7 \\ \hline
  time spent by kernels &  24.3 &  49.6 &  78.7 &  119.0 &  176.4 &  259.8 &  332.3 &  431.7 \\
        wall clock time & 111.0 & 343.0 & 626.0 & 2255.0 & 1923.0 & 4269.0 & 3445.0 & 4401.0 \\ \hline \hline
      kernel time flops &  76.4 & 191.5 & 330.6 &  458.3 &  556.7 &  616.1 &  732.2 &  813.1 \\
       wall clock flops &  16.8 &  27.7 &  41.6 &   24.2 &   51.1 &   37.5 &   70.6 &   79.8 \\ \hline \hline
 & \multicolumn{8}{c}{times on the V100} \\
\multicolumn{1}{c||}{stage in Algorithm~1}
 & \multicolumn{1}{c|}{$32$} 
 & \multicolumn{1}{c|}{$64$} 
 & \multicolumn{1}{c|}{$96$} 
 & \multicolumn{1}{c|}{$128$}
 & \multicolumn{1}{c|}{$160$} 
 & \multicolumn{1}{c|}{$192$}
 & \multicolumn{1}{c|}{$224$}
 & \multicolumn{1}{c}{$256$} \\ \hline \hline
 invert diagonal tiles &  1.9 &  11.4 &  21.2 &  36.3 &   61.8 &   78.9 &  103.3 &  138.2 \\
multiply with inverses &  6.4 &  12.7 &  18.2 &  23.9 &   38.9 &   47.1 &   55.2 &   63.1 \\
     back substitution & 11.3 &  13.8 &  19.8 &  26.2 &   44.2 &   58.6 &   78.6 &  113.2 \\ \hline
 time spent by kernels & 19.6 &  37.8 &  59.2 &  86.4 &  145.0 &  184.6 &  237.1 &  314.5 \\
       wall clock time & 90.0 & 251.0 & 482.0 & 776.0 & 1181.0 & 1577.0 & 2150.0 & 2886.0 \\ \hline \hline
     kernel time flops & 94.9 & 250.9 & 439.6 & 631.7 &  677.4 &  867.0 & 1025.9 & 1115.9 \\
      wall clock flops & 20.7 &  37.8 &  54.0 &  70.3 &   83.1 &  101.5 &  113.2 &  121.6
\end{tabular}
\caption{Tiled accelerated back substitution in quad double precision
on the RTX 2080, the P100, and the V100.  
The dimension of the matrices are multiples of 80,
that is: $80 \times n$, where $n = 32, 64, 96, 128, 160, 192, 224$, and~256.}
\label{tabP100V100dbl4tabs}
\end{center}
\end{table}

\subsection{Tiled Back Substitution in Quad Double Precision}

Table~\ref{tabV100nofixNtabs4} lists times for different
choices of $N$ and $n$.
The V100 has 80 streaming multiprocessors,
so in Table~\ref{tabP100V100dbl4tabs}, $N = 80$ and multiples of~32
are taken for $n$, in runs on matrices of dimension 2,560, 5,120, 7,680, 
10,240, 12,800, 15,360, 17,920, and 20,480.
Teraflop performance on the V100 is attained for $n = 224$,
at dimension~$80 \times n = 17,920$.
Reading the first two lines of Table~\ref{tabP100V100dbl4tabs},
observe that the time to invert the diagonal tiles increases
from a tiny 1.9 to 11.4 milliseconds as the dimension doubles,
and from $n = 96$ on, the time spent on inverting the diagonal
tiles dominates the times of the other two stages.
The difference between the wall clock time
and the time spent by all kernels is significant.

\begin{figure}[hbt]
\begin{center}
{\includegraphics[width=9cm]{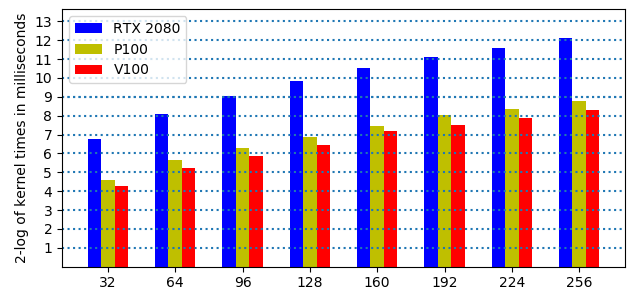}}
\caption{2-logarithms of the times spent by all kernels for
the back substitution on the RTX~2080, the P100, and the V100 
in quad double precision.}
\label{figdbl4tabs}
\end{center}
\end{figure}

Figure~\ref{figdbl4tabs} shows the evolution of the times
spent by all kernels listed in Table~\ref{tabP100V100dbl4tabs}.
In the 2-logarithm plot, an increase of one unit in the height of a bar
equals a doubling of the time. 
For which dimensions is the cost of Algorithm~1 proportional to~$Nm$?
If the dimension doubles from 2,560 to 5,120 and from 5,120 to 10,240
(corresponding to the bars for 32, 64, and 128 
in Figure~\ref{figdbl4tabs}), the doubling of the time is observed
for the P100 and the V100, for the RTX 2080, the increase from
dimension 5,120 to 10,240 is more than three times.

Computing the ratios of the times spent by all kernels
on the P100 over the V100 gives $732.2/237.1 \approx 3.1$ for dimension 17,920
and $813.1/314.5 \approx 2.6$ for dimension 20,480.
That those ratios are still far above the expected 1.68 ratio
is most likely because the number 80 (of blocks and tiles)
coincides with the number of streaming multiprocessors of the V100,
whereas the P100 has 64 streaming multiprocessors.

What is the best choice of $N$ and $n$ 
for a matrix of dimension~20,480?
For the parallelism in GPU acceleration, 
fixing $N$ at 80 gives the best performance as illustrated
in Table~\ref{tabV100nofixNtabs4}.
Doubling $n$ from 64 to 128, and to 256 increases the time
spent by all kernels, but decreases the total wall clock
time from 2.620 seconds to 2.071 seconds,
as the performance then nearly doubles.

\begin{figure}[hbt]
\begin{center}
{\includegraphics[width=9cm]{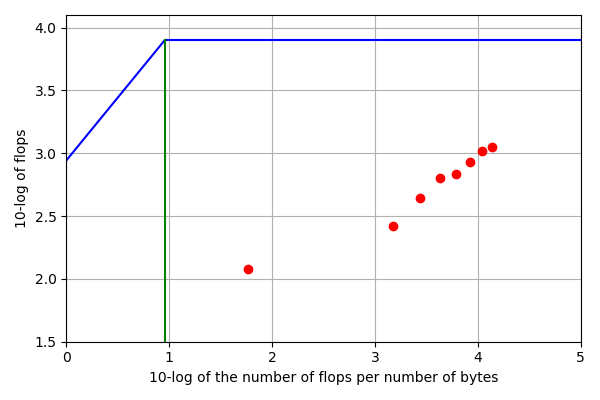}}
\caption{Roofline plot for the data in Table~\ref{tabV100tabs4roofline}.
The first coordinate of each dot is the 10-log of the arithmetic
intensity and the 10-log of the flops is the second coordinate of
each dot.  As $n$ increases, the dots move upwards to the right,
illustrating that the problem becomes more compute bound.  }
\label{figdbl4tabsroofline}
\end{center}
\end{figure}

The roofline model~\cite{WWP09} is applied to visualize the performance.
The {\em arithmetic intensity} of a computation is the ratio of the
number of floating point operations over the number of bytes
in the computation.  
For the V100, the ridge point is computed as 7900/870 = 9.08,
as the ratio of the theoretical peak performance and the memory bandwidth.
Problems with an arithmetic intensity larger than~9 are compute bound, 
as their performance is bounded by the theoretical peak performance 
of 7.9 TFLOPS.  A problem with an arithmetic intensity less than~9
is memory bound, as its performance is bounded by the memory bandwidth
of 870 GB/second.  
Table~\ref{tabV100tabs4roofline} lists the arithmetic intensities
for the back substitution in quad double precision,
for dimensions that are multiples of~80.
Figure~\ref{figdbl4tabsroofline} shows the roofline model
for this experiment.

\begin{table}[htb]
\begin{center}
\begin{tabular}{r||r|r|r|r|r|r|r|r}
\multicolumn{1}{c||}{~} 
 & \multicolumn{1}{c|}{$32$} 
 & \multicolumn{1}{c|}{$64$} 
 & \multicolumn{1}{c|}{$96$} 
 & \multicolumn{1}{c|}{$128$}
 & \multicolumn{1}{c|}{$160$} 
 & \multicolumn{1}{c|}{$192$}
 & \multicolumn{1}{c|}{$224$}
 & \multicolumn{1}{c}{$256$} \\ \hline
  (1) 
 & \!58.71\! & \!1500\! & \!2740\! & \!4308\! & \!6203\! 
 & \!8427\! & \!10980\! & \!13860 \\
  (2) 
 & \!119.1\! & \!263.9\! &  \!440.7\! &  \!633.8\! &  \!679.0\!
 & \!852.9\! & \!1036.0\! & \!1113.6 
\end{tabular}
\caption{Arithmetic intensity (1) and the kernel time flops (2)
for the tiled accelerated back substitution in quad double precision
on the V100.  The dimensions are multiples of 80, that is: 
$80 \times n$, where $n = 32, 64, 96, 128, 160, 192, 224$, and~256.}
\label{tabV100tabs4roofline}
\end{center}
\end{table}

The leftmost dot in Figure~\ref{figdbl4tabsroofline}
is an outlier because at $n=32$, the V100 is only half occupied,
as the V100 has 64 cores per streaming multiprocessor.
For an increasing number of threads per block,
the arithmetic intensity increases.

\subsection{Least Squares Solving in Four Different Precisions}

\begin{table}[htb]
\begin{center}
\begin{tabular}{r||r|r|r|r}
  & \multicolumn{4}{c}{times on the RTX 2080} \\
  \multicolumn{1}{c||}{stage}
                   & \multicolumn{1}{c|}{1d} 
                   & \multicolumn{1}{c|}{2d}
                   & \multicolumn{1}{c|}{4d}
                   & \multicolumn{1}{c}{8d} \\ \hline \hline
    QR kernel time & 327.4 & 4082.2 & 36128.9 & 164626.8 \\ 
      QR wall time & 565.0 & 4785.0 & 37347.0 & 167002.0 \\ \hline
    BS kernel time &   1.7 &   20.8 &   192.0 &    895.1 \\
      BS wall time &   4.0 &   26.0 &   200.0 &    910.0 \\ \hline \hline
   QR kernel flops & 146.3 &  252.2 &   281.7 &    292.7 \\
     QR wall flops &  85.0 &  215.2 &   272.6 &    288.6 \\ \hline
   BS kernel flops &   9.7 &   17.3 &    18.7 &     19.1 \\
     BS wall flops &   4.1 &   13.9 &    17.9 &     18.8 \\ \hline \hline
total kernel flops & 145.6 &  251.0 &   280.3 &    291.3 \\
  total wall flops &  84.2 &  214.1 &   271.2 &    287.1 \\ \hline \hline
  & \multicolumn{4}{c}{times on the P100} \\
  \multicolumn{1}{c||}{stage}
                   & \multicolumn{1}{c|}{1d} 
                   & \multicolumn{1}{c|}{2d}
                   & \multicolumn{1}{c|}{4d}
                   & \multicolumn{1}{c}{8d} \\ \hline \hline
    QR kernel time & 268.9 &  707.8 & 5193.0 & 20508.2 \\ 
      QR wall time & 319.0 &  822.0 & 5373.0 & 20853.0 \\ \hline
    BS kernel time &   4.0 &    7.5 &   40.8 &   181.8 \\
      BS wall time &   6.0 &   11.0 &   48.0 &   200.0 \\ \hline \hline
   QR kernel flops & 178.2 & 1454.7 & 1960.1 &  2349.9 \\
     QR wall flops & 150.2 & 1252.5 & 1894.3 &  2311.0 \\ \hline
   BS kernel flops &   4.1 &   47.8 &   87.8 &    94.0 \\
     BS wall flops &   2.9 &   32.4 &   74.1 &    85.4 \\ \hline \hline
total kernel flops & 175.6 & 1439.9 & 1945.5 &  2330.1 \\
  total wall flops & 147.6 & 1236.2 & 1878.1 &  2289.9 \\ \hline \hline
  & \multicolumn{4}{c}{times on the V100} \\
  \multicolumn{1}{c||}{stage}
                   & \multicolumn{1}{c|}{1d} 
                   & \multicolumn{1}{c|}{2d}
                   & \multicolumn{1}{c|}{4d}
                   & \multicolumn{1}{c}{8d} \\ \hline \hline
    QR kernel time & 157.9 &  451.1 & 3020.6 & 11924.5 \\ 
      QR wall time & 204.0 &  566.0 & 3203.0 & 12244.0 \\ \hline
    BS kernel time &   2.0 &    4.0 &   28.0 &   114.5 \\
      BS wall time &   4.0 &    7.0 &   35.0 &   127.0 \\ \hline \hline
   QR kernel flops & 303.4 & 2282.2 & 3369.8 &  4041.4 \\
     QR wall flops & 235.1 & 1819.6 & 3177.8 &  3936.1 \\ \hline
   BS kernel flops &   8.1 &   89.8 &  127.9 &   149.1 \\
     BS wall flops &   4.2 &   49.8 &  102.9 &   134.5 \\ \hline \hline
total kernel flops & 299.6 & 2262.9 & 3340.0 &  4004.4 \\
  total wall flops & 230.8 & 1797.3 & 3144.7 &  3897.0
\end{tabular}
\caption{Least squares solving in double (1d), double double (2d),
quad double (4d), and octo double (8d) precision,
on a 1,024-by-1,024 linear system, with 8 tiles of size 128,
on the RTX 2080, the P100, and the V100.  BS = Back Substitution.}
\label{tabqrbs4prc}
\end{center}
\end{table}

Table~\ref{tabqrbs4prc} summarizes the times and the flops
of solving a linear system in the least squares sense,
in four different precisions.
The blocked accelerated Householder QR of Algorithm~2
is followed by Algorithm~1, the tiled accelerated back substitution.

Comparing the kernel times in Table~\ref{tabqrbs4prc}
in all precisions shows that the time for the back substitution
is about 100 times less than the time for the QR decomposition.
Consequently, the lower performance of the back substitution
in small dimensions does not lead to a significant reduction
in the overall performance of the solver.

As the QR decomposition has a cost that is cubic in the dimension,
versus the quadratic cost of the back substitution,
one could have expected at dimension 1,024 to see a factor of one
thousand in the ratios between the QR and the back substitution.
Or equivalently, the times for the QR would have been one thousand
times longer than for the back substitution.  
Instead, the observed factor is closer to one hundred than one thousand,
thanks to the well performing GPU accelerated QR.

As a final observation, times on the QR decomposition
of a random 1,024-by-1,024 matrix in quad double precision,
on the V100 appear in 
Table~\ref{tabqr4prc}, Table~\ref{tabV100baqr4d}, Table~\ref{tabqrbs4prc},
with respective kernel times 3167.0, 3136.5, 3020.6, and
respective wall clock times 3356.0, 3366.0, 3203.0,
illustrating the fluctuations of the measured milliseconds.

\section{Conclusions}

Taking 439, the average number of double operations in the tallies
of the operational counts for quad double arithmetic,
as the scaling factor, teraflop performance on a GPU can be viewed
as 2.2 gigaflops on a single threaded computation.
Using this interpretation, the experiments show 
that GPU acceleration does compensate the overhead cost
of quad double arithmetic.
In any case, the observed cost overhead ratios in going from 
double double to quad double are less than the ratios predicted
by the operational count tallies.


\end{document}